\documentclass{llncs}

\usepackage{times,amsmath,amssymb,epsfig}
\usepackage{url}
\usepackage[colorlinks=true,
           urlcolor=blue,
           citecolor=blue,
           linkcolor=blue,
          bookmarks=false,
           bookmarksnumbered,
           linktocpage=true
           ]{hyperref}
           
           \usepackage{color}
          
           \definecolor{comment}{rgb}{0.4,0.6,0.9}
           \definecolor{fixme}{rgb}{1,0.3,0.3}

\newcommand{\pol}{\mathit{pol}}

\newcommand{\pg}{\mathit{PG}}
\newcommand{\en}{\mathit{En}}
\newcommand{\act}{\mathit{Act}}
\newcommand{\ag}{\mathit{Ag}}

\newcommand{\used}{\mathit{used}}
\newcommand{\wgby}{\mathit{genBy}}

\newcommand{\wdf}{\mathit{wasDerivedFrom}}
\newcommand{\waw}{\mathit{waw}}
\newcommand{\attrTo}{\mathit{wat}}

\newcommand{\delegate}{\mathit{abo}}

\newcommand{\preorder}{\preceq}

\newcommand{\type}{\mathit{type}}

\newcommand{\guEA}{\pg_{gu/ea}}  

\newcommand{\group}{\mathit{Group}}

\newcommand{\clos}{\mathit{pclos}}
\newcommand{\repl}{\mathit{replace}}

\newcommand{\extend}{\mathit{extend}}

\newcommand{\aStart}{\mathit{startEv}}
\newcommand{\aEnd}{\mathit{endEv}}
\newcommand{\genEv}{\mathit{genEv}}
\newcommand{\useEv}{\mathit{useEv}}

\newcommand{\sens}{\ensuremath{s}}

\newcommand{\provabs}{\texttt{ProvAbs}}

\newcommand{\source}[1]{\ensuremath{\mathit{source}(#1)}}

\title{\textit{ProvAbs}: model, policy, and tooling for abstracting PROV graphs\thanks{This work was funded in part by EPSRC UK and DSTL under grant EP/J020494/1}}

\author{
Paolo Missier\inst{1} \and Jeremy Bryans\inst{1}  \and Carl Gamble\inst{1} \and Vasa Curcin\inst{2} \and Roxana Danger\inst{2}}
\institute{School of Computing Science, Newcastle University \and
Imperial College, London}

\begin{document}
\maketitle
\begin{abstract} 
Provenance metadata can be valuable in data sharing settings, where it can be used to help data consumers form judgements regarding the reliability of the data produced by third parties.
However, some parts of provenance may be sensitive, requiring access control, or they may need to be simplified for the intended audience.
Both these issues can be addressed by a single mechanism for creating abstractions over provenance, coupled with a policy model to drive the abstraction. 
Such mechanism, which we refer to as \textit{abstraction by grouping}, simultaneously achieves partial disclosure of provenance, and facilitates its consumption.
In this paper we introduce a formal foundation for this type of abstraction, grounded in the W3C PROV model; describe the associated policy model; and briefly present its implementation, the \provabs{} tool for interactive experimentation with policies and abstractions.
\end{abstract}

\section{Introduction}\label{sec:introduction}

Provenance, a formal representation of the production process of data, may facilitate the assessment and improvement of the quality of data products, as well as the validation and reproducibility of scientific experimental datasets.
This expectation predicates on an  assumption of interoperability between mutually independent producers and consumers of provenance. 
The W3C PROV generic provenance model \cite{w3c-prov-dm} is intended to facilitate such interoperability, by providing a common syntax and semantics for provenance models, and thus enable provenance-aware data sharing at Web scale. 
%

 
\vspace{-10pt}
 \subsection{Abstracting provenance}

For provenance to be useful, it must be represented at a level of abstraction that is appropriate to the consumer. For example, system-level provenance which includes individual system calls and I/O operations may be appropriate for system auditing purposes, while a higher level description may be more appropriate to determine how a document evolved to its final version, e.g. through a series of edits involving multiple authors.
In some cases, the higher abstraction can be computed from the detailed representation. 
One such case occurs when provenance describes the execution of a workflow or dataflow, which can itself be described at multiple levels of abstraction.
Early work on \textit{provenance views} (\textit{Zoom}) \cite{DBLP:conf/icde/BitonBDH08} is an example. Here users specify the abstraction they require on the workflow, and that is used to compute a corresponding abstract view of the workflow's trace.
More generally, however, a trace may represent arbitrary process executions and data derivations, and one cannot rely on a formal description of the process to specify a suitable abstraction.

The problem of abstracting over provenance in such a more general setting has been addressed in later work, notably the ProPub system \cite{springerlink:10.1007/978-3-642-22351-8_13}. Here the main goal is to ensure that sensitive elements of the trace are abstracted out, by means of a redaction process.
In ProPub, users specify edit operations on a provenance graph, such as anonymizing, abstracting, and hiding certain parts of it.
ProPub operates on a simplified provenance model (which pre-dates PROV) which only includes use/generation relations, and adopts an ``apply--detect--repair'' approach. First, user-defined abstraction rules are applied to the graph, then consistency violations that may occur in the resulting new graph are detected, and  finally a  set of edits are applied to repair such violations. In some cases, this causes nodes that the user wanted removed to be reintroduced, and it is not always possible to satisfy all user rules. 

\vspace{-10pt}
\subsection{Contributions}

 Our work is motivated by the need to control the complexity of a provenance graph by increasing its level of abstraction, as well as to protect the confidentiality of parts of the graph.
 Our specific contributions in this paper are threefold.
 Firstly, we define a \textit{Provenance Abstraction Model} (\textbf{PAM})  centred on the $\group$ abstraction operator. $\group$ replaces a set of nodes $V_{gr} \subset V$ in a valid PROV graph $\pg$ with a new abstract node, resulting in the modified graph $PG'$.
 The rewriting  preserves the validity of the graph, in the sense made precise below, and it does not introduce any new relations into $\pg'$, which are not justified by existing $\pg$ relations.
A formal account of this operator is given in Sec.\ref{sec:grouping}.
A preliminary but more extended account of this work appears in our technical report \cite{w3c-prov-dm}.

Secondly, we present a simple policy model and language for controlling abstraction, based on the assumption that provenance \textit{owners} want to control the disclosure of their provenance graphs to one or more \textit{receivers}, with varying levels of trust (Sec.\ref{sec:policy}).
The model lets the owners associate a policy, $\pol$, to a graph. 
Policy evaluation results in a \textit{sensitivity} value $\sens(v,\pol)$ being associated to each node $v$.
Assuming, as in the Bell-Lapadula model \cite{bell1996bell}, that a \textit{clearance level} $cl$ can be associated to each receiver, the nodes $V_{gr}$ to be abstracted in $\pg$ according to $\pol$ are those for which 
$\sens(v,\pol) > cl$.

Finally, we present the \provabs{} tool, which implements both $\group$ and the policy language. \provabs{} has been demonstrated on our confidentiality preservation use case, in the context of intelligence information exchange \cite{Missier2013d}.
 
 \vspace{-10pt}
\subsection{Related work}

In addition to the Zoom and ProPub prototypes cited above, strands of research that are relevant to this work include (i) provenance-specific graph redaction, (ii) graph anonymization, and (ii) Provenance Access Control (PAC).
Provenance redaction \cite{Cadenhead:2011:TPU:1998441.1998456} employs a graph grammar technique to edit provenance that is expressed using the Open Provenance Model \cite{Moreau2010a}(a precursor to PROV), as well as a redaction policy language. The critical issue of ensuring that specific relationships are preserved, however, is addressed only informally in the paper, i.e., with no reference to OPM semantics.

Extensions to the relational data anynomization framework to graph data structures, specifically for social network data, have been developed \cite{springerlink:10.1007/978-3-540-78478-4_9,Bhagat:2009:CGA:1687627.1687714,Liu:2008:TIA:1376616.1376629}.
The approach, involving randomly removing and adding arcs, will not work for PROV, however, as it would result in new, false dependencies.
More relevantly, PAC is concerned with enforcing access control on parts of a provenance graph, in the context of secure provenance exchange. 
An analysis of the associated challenges \cite{Braun:2008:SP:1496671.1496675} notes that provenance of data can be more sensitive than the data itself.
In a similar setting, \cite{Hasan:2007:ISP:1314313.1314318} accounts for the possibility of forgery of provenance by malicious users, and of collusion amongst users to reveal sensitive provenance to others. 
However, the paper stops short of providing any hints at technical solutions, and indeed it is not clear how these problems are specific to provenance, as opposed to data sharing in general.
Finally, our policy language is loosely related to an XACML-based policy language \cite{Cadenhead:2011:LPA:1943513.1943532} the access control system for provenance, where path queries are used to specify target elements of the graph.

 \vspace{-10pt} 
 \section{Essential PROV}
 
 We now introduce the PROV concepts that are required for the rest of the paper.
The PROV data model\cite{w3c-prov-dm} defines three types of sets: (i) Entities ($\en$),  i.e., data, documents; (ii) Activities ($\act$), which represent the execution of some process over a period of time, and (iii) Agents ($\ag$), i.e., humans, computing systems, software. The following set of core relations is also defined amongst these sets:
\small
\begin{align*}
\text{usage:} & \used  \subseteq \act \times \en  & 
\text{generation:} & \wgby \subseteq  \en \times \act \\
\text{derivation:} & \wdf \subseteq  \en \times \en  &
\text{association:} &  \waw \subseteq \act \times \ag \\
\text{delegation:} &  \delegate \subseteq \ag \times \ag &
\text{attribution:} & \attrTo \subseteq \en \times \ag
\label{eq:prov-relations}
\end{align*}
\normalsize

For simplicity and due to space constraints, in this paper we restrict our scope to just  $\en$, $\act$, and relations $\used$ and $\wgby$. The extension of this work to Agents and their relations ($\delegate$, $\attrTo$), is available from our extended tech report \cite{Missier2013d}.
The extension to other core relations such as $\wdf$ is straightforward and will not be discussed here.

We denote instances of these relations as  $\wgby(e,a)$, $\used(a,e)$, etc., where  $e \in \en, a \in \act$.
Following common practice, we view a set $I$ of such binary relation instances as a digraph $G=(V,E)$, where $V= \en \cup \act$ and $E$ is a set of labelled edges, and where $x \xleftarrow{r} y \in E$ iff $r(x,y) \in I.$\footnote{Conventionally, we orient these edges from right to left, to denote that the relation ``points back to the past''.}
Finally, we denote the set of all such provenance graphs as $\guEA$, to indicate that they only contain $\wgby$ and $\used$ relations amongst $\en$ and $\act$ nodes.

%
\begin{figure*}[tb]
\begin{center}
\includegraphics[width=.85\textwidth]{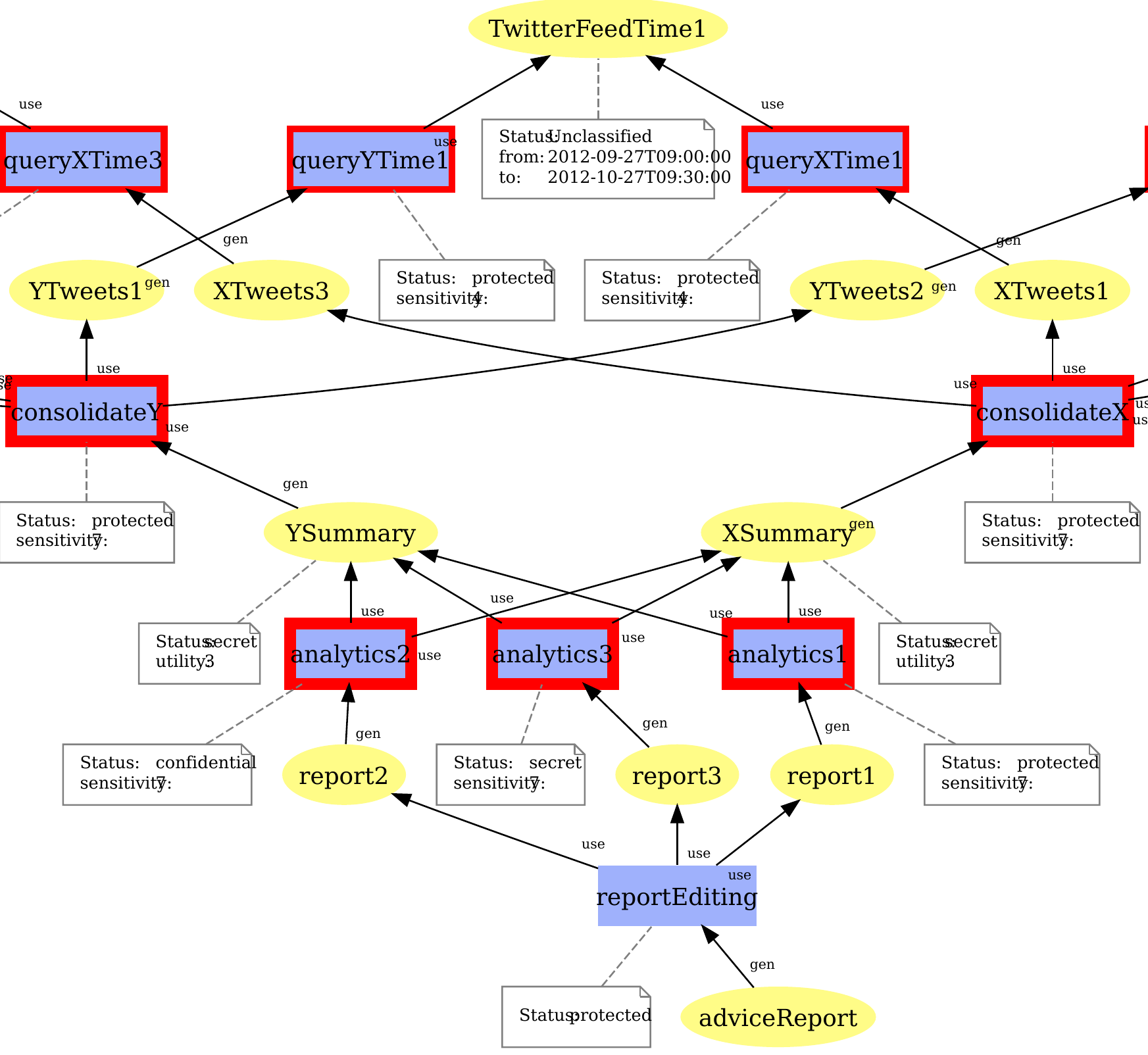}
\caption{Example provenance graph of a complex document production process. The  \provabs{} model is designed to abstract some of the elements in the graph, for instance to avoid their disclosure. Coloured boxes denote \provabs\ sensitivity annotations, explained in Sec.~\ref{sec:policy}.}
\label{fig:sensitivities}
\end{center}
\end{figure*}
Fig.~\ref{fig:sensitivities} shows an example of a $\guEA$ graph, where ovals and rectangles represent Entities, Activities, and Agents, respectively.
The graph describes a document, \texttt{advice-report}, which was ultimately derived from twitter feeds captured at different times, through a series of query, consolidation, and analysis activities.
The agents to whom the documents and activities are ascribed are omitted for simplicity. Note also that the nodes are decorated with user-defined properties, such as \texttt{Status}.

A set of formal constraints are defined on the PROV data model. These are described in the PROV-CONSTRAINTS document \cite{w3c-prov-constraints}. Two groups of constraints are relevant here.
The first (Constraint 50 --- typing\footnote{\url{http://www.w3.org/TR/prov-constraints/\#typing}}) formalises the set-theoretical definitions of the relations given above.
Additionally, Constraint 55\footnote{\url{http://www.w3.org/TR/prov-constraints/\#entity-activity-disjoint}} stipulates that entities and activities are disjoint: 
$  \en \cap \act = \emptyset $.

The second group concerns temporal ordering amongst events. 
PROV defines a set of instantaneous events which mark the lifetime boundaries of Entities (generation, invalidation), Activities (start, end), and Agents (start, end), as well as some of the interactions amongst those elements, such as generation and usage of an entity by an activity, attribution of an entity to an agent, and more. 
Optionally, events may be explicitly associated to PROV elements. 
In the following, we denote the start and end of an activity $a$ by 
$\aStart(a)$, $\aEnd(a)$, respectively, and the generation and usage events for an entity $e$ and activity $a$ with $\genEv(\wgby(e,a))$, $\useEv(\used(a,e))$, respectively (as mentioned, Agents are beyond the scope of this paper).
PROV events form a preorder, which we denote $\preorder$.
The relevant temporal constraints are expressed as follows.

\begin{itemize}
\item \textbf{C1: generation-generation-ordering (Constraint 39):}  If an entity is generated by more than one activity, then the generation events must all be simultaneous. Let $e \in \en, a_1, a_2 \in \act$, and let $\wgby(e, a_1)$ and $\wgby(e, a_2)$ hold. Then the following must hold:
\vspace{-10pt}
\begin{align*}
\genEv(\wgby(e, a_1)) &\preorder \genEv(\wgby(e, a_2)) \text{  and} \\
\genEv(\wgby(e, a_2)) & \preorder \genEv(\wgby(e, a_1))
\end{align*}

\item\textbf{C2: generation-precedes-usage(Constraint 37):} A generation event for an entity must precede any usage event for that entity.
Let $a \in \act$, $e \in \en$, and let $\used(a,e))$, $\wgby(e,a)$ hold. Then:
\[\genEv(\wgby(e, a)) \preorder \useEv(\used(a,e))\]

\item\textbf{C3: usage-within-activity (Constraint 33):} Any usage of $e \in \en$ by some $a \in \act$ cannot precede the start of $a$ and must precede the end of $a$. Let $\used(a,e)$ hold. Then:
\[\aStart(a) \preorder \useEv(\used(a,e))   \preorder \aEnd(a)\]

\item\textbf{C4: generation-within-activity (Constraint 34):} The generation of $e$ by $a$ cannot precede the start of $a$ and must precede the end of $a$.
If $\wgby(e,a)$, then:
\[ \aStart(a) \preorder \genEv(\wgby(e,a))  \preorder \aEnd(a)\]

\end{itemize}

A \textit{valid} PROV graph is one that satisfies all the constraints defined in the PROV-CONSTR document~\cite{w3c-prov-constraints}.
Within our scope, a valid $\guEA$ graph is one that satisfies the constraints defined here.

 \vspace{-10pt}
 \section{Abstraction by grouping}  \label{sec:grouping}

Simple edits that can be applied to a graph to protect confidentiality of its content include removing individual nodes or edges. 
Alternatively, the node's identity can be changed, or the values associated to any of its properties can be removed.
These straightforward edits are legal in PROV and they will not be discussed further.
\footnote{Note that removing an arbitrary node may result in disconnected fragments of the graph, as in general one cannot simply add edges to reconnect the remaining nodes, unless those can be inferred from standard PROV constraints. For instance, if activity $a$ is removed from the graph: $\{\used(a, e_1)$, $\wgby(e_2, a) \}$, this results in two disconnected nodes $e_1$, $e_2$, because no relationship can be inferred between them from the original graph.}
We are instead concerned with edits that replace a group of nodes with a new abstract node.

\subsection{Core concepts}

To model this type of abstraction, we are going to define a $\group$ operator which 
takes a graph $G = (V, E) \in \guEA$ and a subset $V_{gr} \subset V$ of its nodes, and produces a modified graph $G' = (V',E') \in \guEA$, where $V_{gr}$ is replaced with a new single node. 
$\group$ is closed under composition, thus allowing for further abstraction by repeated grouping (abstraction of abstraction).
Let $v_{abs} \in V'$ be an abstract node in $G'$. We denote the set $V_{gr}$ of nodes in $G$ that it replaces by $\source{v_{abs}}$.

\begin{figure*}
\begin{center}
\includegraphics[width=.85\textwidth]{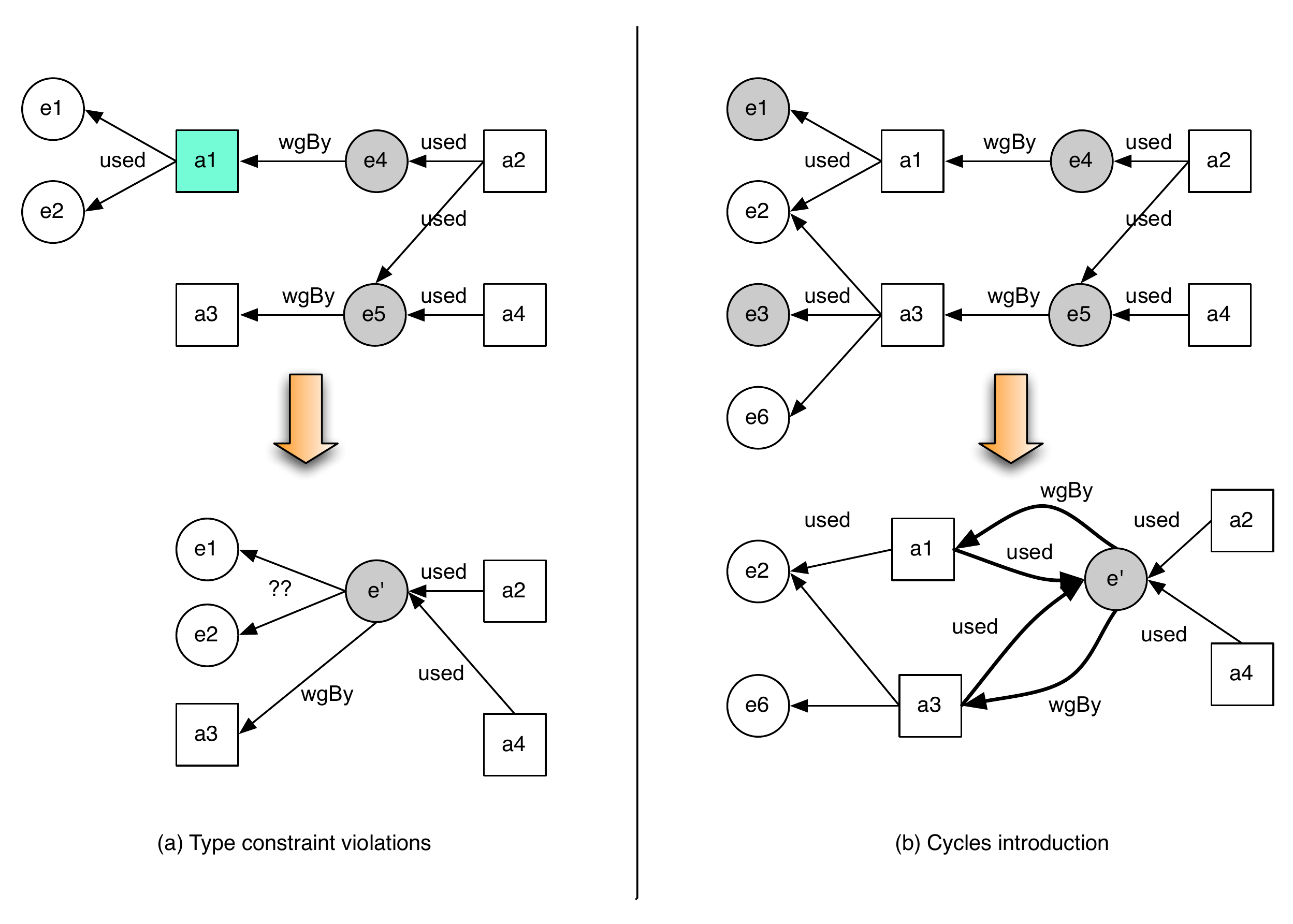}
\caption{Issues with naive replacement of groups of nodes.}
\label{fig:grouping-naive}
\end{center}
\end{figure*}

In order to understand the requirements for defining $\group$, consider the replacements in Fig.~\ref{fig:grouping-naive}.
On the left, nodes $V_{gr} = \{a_1, e_4, e_5\}$ are replaced with a new node $e'$. 
Simply using the original edges to connect the remaining nodes to $e'$ leads to type constraint violations,  namely for the new edges $e_1 \leftarrow e'$, $e_2 \leftarrow e'$, and thus to an invalid graph.

Now consider Fig.\ref{fig:grouping-naive}(b), where $V_{gr} = \{e_1, e_3, e_4, e_5\}$. 
In this case, the simple strategy or replacing $V_{gr}$ with $e'$ and reconnecting the remaining nodes leads to the two cycles: $\{  \wgby(e', a_1), \used(a_1, e')\}$ and 
$\{  \wgby(e', a_3), \used(a_3, e')\}$.
Such cycles are legal, and in particular they are consistent with temporal constraints C1-C4 above. 
Indeed, it is easy to imagine a situation where an activity $a$ first generates an entity $e$, and then makes use of $e$.
For instance, $a$ could be a programming artifact, i.e., an object that first instantiates a new object $e$, and then makes use of $e$.
In this case, the event ordering is 
\begin{equation}
\aStart(a) \preorder \genEv(e,a) \preorder \useEv(a,e) \preorder \aEnd(a)
\label{eq:ordering}
\end{equation}
Yet, we argue that \textit{introducing} new cycles during abstraction is undesirable. 
Intuitively, this is because cycles make stronger assumptions on the possible temporal ordering of events than those in the original graph, and thus are only representative of a restrictive class of graphs. 
To elaborate more precisely on this point, we first introduce new definitions of generation and usage events for an abstract node $v_{abs}$, from the corresponding events associated to $\source{v_{abs}}$.
For this, consider the definition of generation and usage in \cite{w3c-prov-dm}:
\begin{quote}
\textbf{Generation} is the \textit{completion of production} of a new entity (Sec. 5.1.3).\\
\textbf{Usage} is the \textit{beginning of utilizing} an entity  (Sec. 5.1.4).
\end{quote}


An abstract node $v_{abs}$ can be thought of as representing the collection $\source{v_{abs}}$ in the new graph. 
Thus, its ``generation'' is logically defined as the completion of production of its source nodes, that is, its associated generation event should be the \textit{latest} generation event from within its source. 
Note that associating a generation event to an abstract node requires the existence of a generating activity. 
Although this is not always provided as a result of abstraction by grouping, \textit{Inference 7} in \cite{w3c-prov-dm} ensures that such generating activity exists. 
Thus we can formally define generation for abstract nodes, as follows.
\begin{definition}[Abstract node generation event]
Let $V_{gr} \in V$ and $v_{abs}$ be a new abstract node, with $\source{v_{abs}} = V_{gr}$ and generating activity $a$.
Define:
\[ \genEv(\wgby(v_{abs}, a)) = \max_{ e_i \in \source{v_{abs}}}  \genEv(\wgby(e_i,a_i)) \]
where 
$a_i$ is the generating activity of $e_i$.
\label{def:abstract-gen}
\end{definition}
Symmetrically, we associate a usage event to $v_{abs}$, which is the \textit{earliest} usage event for the nodes in $e_i \in \source{v_{abs}}$.
\begin{definition}[Abstract node usage events]
\label{def:abstract-use}
Let $V_{gr} \in V$, $G' = (V',E')$ be the new abstract graph, and let $v_{abs} \in V'$ be a new abstract node.
If there exists an activity $a \in V'$ such that $\used(a, v_{abs})$ holds, then 
\[ \useEv(\used(a, v_{abs})) = \min_{ e_i \in \source{v_{abs}}}  \useEv(\used(a_i,e_i)) \]
where 
$a_i$ is an activity that used $e_i$.
\end{definition}

With these definitions in place, temporal constraint (\ref{eq:ordering}), which applies to simple usage-generation cycles in the graph, translates into the requirement that \textit{every} entity $e_i \in \source{v_{abs}}$ be generated before \textit{any} use of $e_i$.
This constraint ties to each other the generation and usage time of the nodes that are abstracted. In the original graph, however, there is no such requirement: the generation of any entity is, in general, independent of that of others. 
This suggests that  a new generation-usage cycle in the abstract graph adds constraints that are not present in the original graph, and should therefore be avoided.
Note that ProPub \cite{springerlink:10.1007/978-3-642-22351-8_13} also insists on avoiding cycles, but the formal argument in support of this requirement does not appear to be clearly grounded in semantics.

To summarize, the requirements for $\group$ when $G$ is rewritten into $G'$ are: (i) no type constraint violations must occur in $G'$, (ii) no new relationships that are not also present in $G$ are introduced in $G'$, and (iii) no new usage-generation cycles are introduced in $G'$.

\subsection{Convexity, Closure, extensions, and replacement}
\label{sec:closure}

Intuitively, the reason for cycles such as the one in Fig.\ref{fig:grouping-naive}(b) is that set $V_{gr}$ is not ``convex'', that is, there are paths in $G$ that lead out of $V_{gr}$ and then back in again. 
This observation suggests the introduction of a preliminary \textit{closure} operation, aimed at ensuring ``convexity'' and therefore acyclicity. 
This is defined as follows.

\begin{definition}[Path Closure]
\label{def:clos}
Let $G = (V,E) \in \guEA$ be a provenance graph, and let $V_{gr} \subset V$.  
For each pair  $v_i, v_j \in V_{gr}$ such that there is a directed path $v_i \leadsto v_j$ in $G$, let $V_{ij} \subset V$ be the set of all nodes in the path.
The Path Closure of $V_{gr}$ in $G$ is
\[\clos(V_{gr}, V)  =  \bigcup_{v_i, v_j \in V_{gr}} V_{ij} \]
\end{definition}

\begin{figure*}
\begin{center}
\includegraphics[width=\textwidth]{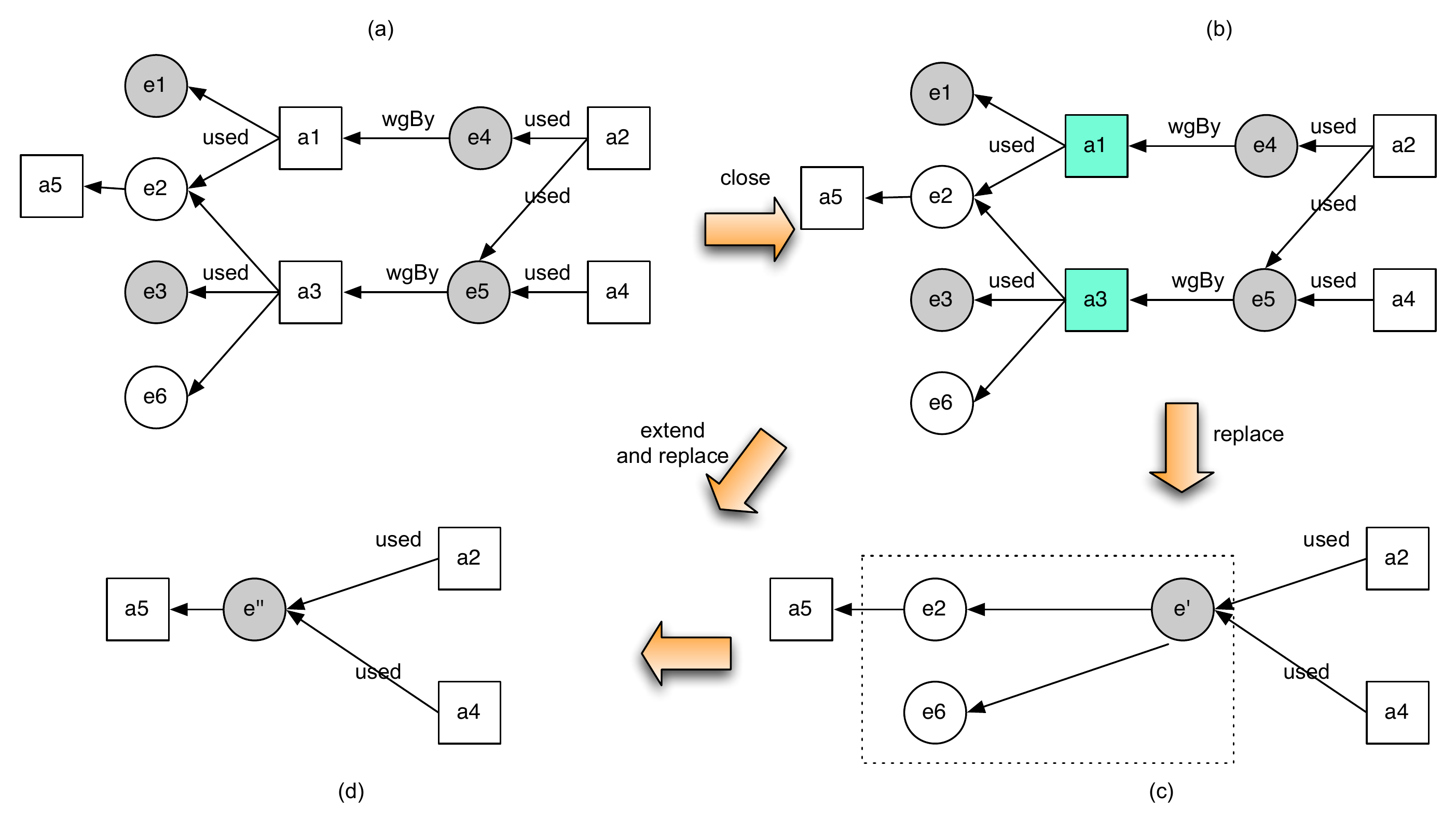}
\caption{Grouping by closure and extension.}
\label{fig:grouping-correct}
\end{center}
\end{figure*}

\sloppy Fig.~\ref{fig:grouping-correct}(b) shows closure applied to the example of Fig.\ref{fig:grouping-naive}, i.e. $\clos(\{e_1, e_3, e_4, e_5\},G) = \{e_1, e_3, e_4, e_5, a_1, a_3\}$. The result of replacing this set with $e'$ is shown in (c).
However, while this solves the cycle problem, the graph still violates type constraints, namely on the new edges $e_2 \leftarrow e'$ and $e_6 \leftarrow e'$.
In this example, we can construct a new group of nodes, $\{ e', e_2, e_6\}$, on the graph that results from the first replacement, and replace it with a new node $e''$. The resulting graph (d) is valid.

To preserve validity in the general case, we are going to first extend the closure in (b) to include e-nodes $e_2$, $e_6$, and then replace the resulting set with $e''$ (the ``extend and replace'' arrow from (b) to (d) in the figure).
Following this approach, $\group$ is defined as the composition of three functions: \textit{closure}, defined above, \textit{extension}, and \textit{replacement}, as follows.

The \textit{extension} of a set $V_{gr} \subset V$ relative to type $t \in \{ \en, \act \}$ is $V_{gr}$ augmented with all its adjacent nodes, in either direction, of type $t$. Formally:
\begin{definition}[$\extend$]
Let $G = (V,E) \in \guEA$, $t \in \{ \en, \act \}$.
\begin{align*}
\extend(V_{gr}, G ,t) =  &
\quad    \{ v' | (v, v') \in E \wedge v \in V_{gr} \wedge \type(v') = t) \} \; \cup \\
& \quad   \{ v | (v', v) \in E \wedge v \in V_{gr} \wedge \type(v') = t) \}   \cup \;\; V_{gr} 
\end{align*}
\end{definition}
In our example:
\[\extend(\{e_1, e_3, e_4, e_5, a_1, a_3\}, G, \en) = \{e_1, e_3, e_4, e_5, a_1, a_3, e_2, e_6\}\]
 Note that all sink nodes in $\extend(V_{gr}, G, t)$ are of type $t$ by construction.
\paragraph{Replacement.}
Let $G=(V,E)$, $V_{gr}' \subset V$ be obtained using $\extend$, and let $v_{new}$ be a new node symbol that does not appear in $V$. 
Function $\repl$ replaces $V'$ with $v_{new}$ in $V$, and connects $v_{new}$ to the rest of the graph, as follows.
Let $\vartheta_{out}(V_{gr}')$, $\vartheta_{in}(V_{gr}')$, and $\vartheta_{int}(V_{gr}')$ denote the set of arcs of $G$ leading out of $V_{gr}'$, leading into $V_{gr}'$,
%
Each arc $(v',v) \in \vartheta_{out}(V_{gr}')$ is replaced with a new arc $(v_{new}, v)$, and each arc $(v,v') \in \vartheta_{in}(V_{gr}')$ is replaced with a new arc $(v,v_{new})$, both of the same relation type. 
Arcs in $\vartheta_{int}(V_{gr}')$ are removed along with the nodes in $V_{gr}'$.
Indeed, all sink nodes in $V_{gr}'$ are of type $t$ as noted above, and so is $v_{new}$ by construction. Thus, sink nodes are replaced by a node $v_{new}$ of the same type. Since the arcs have the same type as those they replace, it follows that $\repl$ preserves type correctness.
It is also easy to verify that each new edge in $G'$ can be mapped to an existing edge in $G$ (proof omitted).
%

%
\begin{definition}[Replace]
\label{def:group-replace}
$\repl (V_{gr}, v_{new}, G) = (V', E')$, where:
\begin{eqnarray*}
V' & = & V  \setminus V_{gr}  \cup \{v_{new}\}\\
E' & = & E  \setminus (\vartheta_{out}(V_{gr}) \cup \vartheta_{in}(V_{gr}) \cup \vartheta_{int}(V_{gr}))  	\cup \vartheta_{out}'(V_{gr})  \cup \vartheta_{in}'(V_{gr})
\end{eqnarray*}
\end{definition}

\subsection{\textit{T-grouping}}
\begin{figure*}
\centering
\includegraphics[width=.83\textwidth]{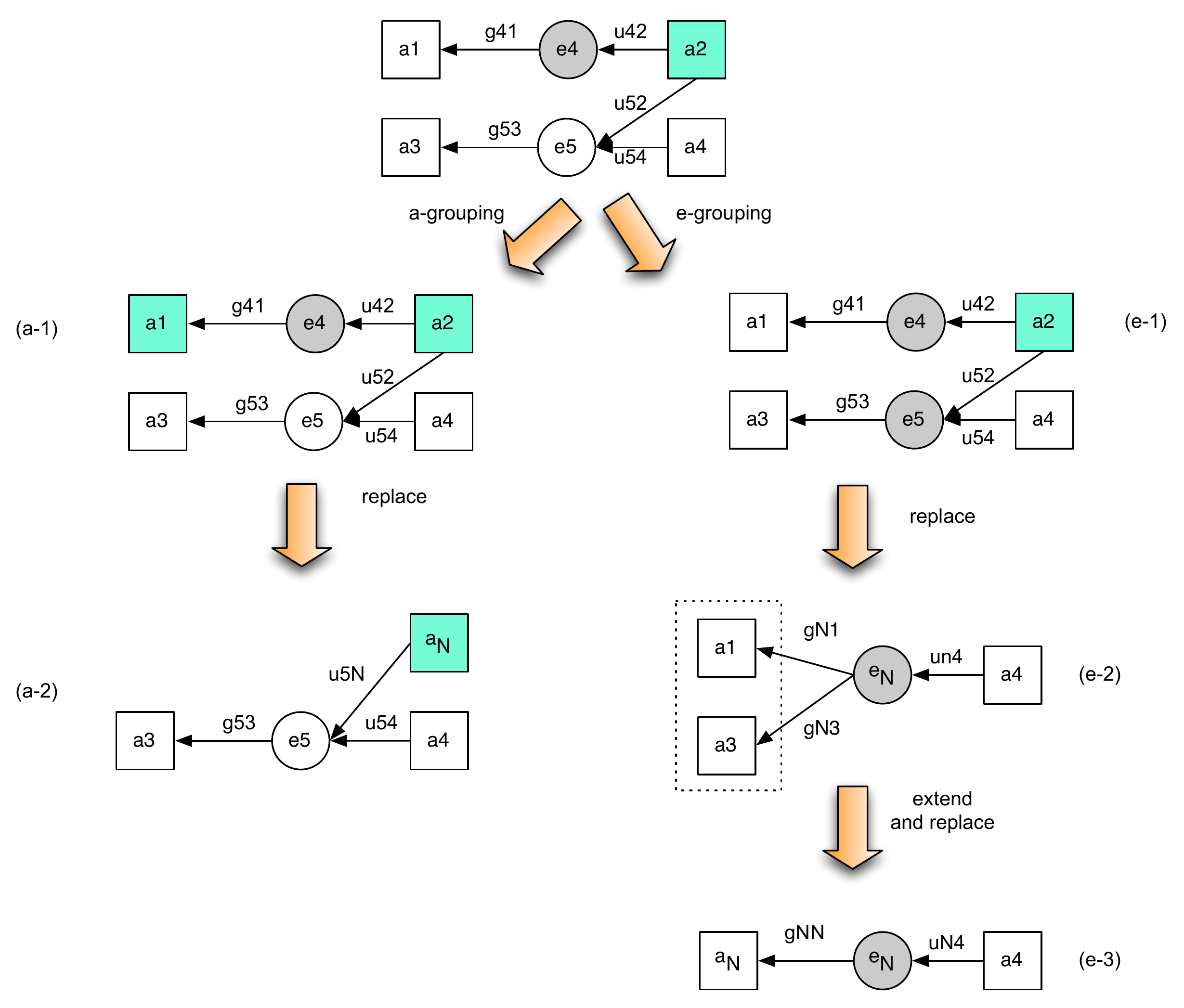} 
\caption{e-grouping and a-grouping} \label{fig:e2-a4}
\end{figure*}

We can now define $\group$ as a composition of closure, extensions, and replacement. 
In general, nodes in $V_{gr}$ can be either $\en$ or $\act$. It is necessary to specify the type of the replacement node, as this may lead to different results.
To make this explicit, we denote the operator by \textbf{t-grouping} (i.e, \textbf{e-grouping} or \textbf{a-grouping}, respectively). 
In the next section, we clarify how user-defined policies are used to control the application of \textbf{t-grouping} to a provenance graph.
\begin{definition}[t-Grouping]
\label{def:t-grouping}
Let $G=(V,E) \in \guEA$, $V_{gr} \in V$, $t \in \{\en, \act\}$, and let  $v_{new}$ be a new node with $\type(v_{new}) = t$.
Then:
\begin{align*} 
\group & (G, V_{gr}, v_{new}, t) = 
 \repl( \extend(\clos(V_{gr},V), V, t), v_{new},  G )
\end{align*}
\label{eq:t-grouping}
\end{definition}
\vspace{-20pt}
Note sink nodes in the closure are homogeneous and are replaced by a node of the same type $t$. 
This satisfies the necessary condition for $\repl$ to perform correctly.
Fig.~\ref{fig:e2-a4}(a-1, a-2) illustrates $\group(G,\{ e_4, a_2\}, v_{new}, \act)$, while 
Fig.~\ref{fig:e2-a4}(e-1, e-2,e-3) shows $\group(G,\{ e_4, a_2\}, v_{new}, \en)$.
Note that a new pattern arises in the case of \textit{e-grouping} as shown in Fig.~\ref{fig:e2-a4}(e-1, e-2). Now the extension leads to $V_{cl} = V_{gr} \cup \{ e_5\}$, which in turn leads to the pattern shown in Fig.~\ref{fig:e2-a4}(e-3), involving two generation events for the new entity $e_{N}$.
Although this is a valid pattern, the two generation events must be simultaneous by C1 above.
The intuitive interpretation for this pattern is that each of the two activities generated one entity in $\source{e_N}$, and that the abstraction makes these two events indistinguishable. Formally, nothing further needs to be done to the graph. However one can restore the more natural pattern whereby one single generation event is recorded for $e_N$, by propagating the grouping to the set of generating activities. In the example, this leads to the graph in Fig.~\ref{fig:e2-a4}(e-3). 

\vspace{-20pt}

 \vspace{-10pt}
 \section{Policy model}  \label{sec:policy}

Having outlined the grouping operator, we now present a simple policy language to let users specify one or more grouping sets $V_{gr}$ for abstraction. We refer to these users as Policy Setters (PS).
%
%
%
%
Our approach consists of two phases.
The first phase involves annotating each node $n$ with a \textit{sensitivity} value  $s(n)$ and/or a \textit{utility} value $u(n)$. 
These annotations are independent of any intended receiver of the abstracted graph. 
In the second phase, a grouping set $V_{gr}$ is generated for a specific receiver $r$, denoted $V_{gr}(r)$ for clarity. 
We assume, as in Bell-Lapadula \cite{bell1996bell}, that a pre-defined clearance level $\mathit{cl}(r)$ is associated with $r$.
The nodes to be abstracted are simply those with sensitivity higher than  $\mathit{cl}(r)$:
 $V_{gr}(r) = \{ v \in V | s(n) \geq \mathit{cl}(r) \}$.
 
A policy is a sequence of rules. Each rule (i) identifies a set of nodes, and (ii) assigns a sensitivity to each of those nodes.
Node selection is achieved using a simple form of path expressions on the graph, combined with filter conditions. 
 Keeping simplicity of use by non-expert PS in mind, we have chosen a simplified fragment of regular path expressions on graphs~\cite{mendelzon1995finding}.
The example rules in Fig. \ref{fig:policy-rules} apply to the graph in Fig.~\ref{fig:sensitivities}:
\begin{figure}
\scriptsize
\vspace*{-15pt}
\ttfamily
\begin{tabbing}
***\= **\= *************\=  \kill \\
list classifications\;\;\;    [Unclassified, Classified, Protected, Secret]; \\
for all (act used data) \\
\> where (data.Status $>=$ Secret in classifications  (def true)) setSensitivity(act, 7); \\
for all (process used data) \\
\>where (data descendantOf d14)) setSensitivity(data, 10);
\end{tabbing}
\normalfont
\normalsize
 \vspace{-15pt}
\caption{Example Policy rules}
\label{fig:policy-rules}
\end{figure}

 \vspace{-15pt}
The rules are executed in sequence. 
\sloppy \texttt{List} declares a domain-specific \textit{ordered} enumeration of constants, called \texttt{Classifications}.
The path expression in the first command is a simple pattern where \texttt{act} and \texttt{data} are variables, and \texttt{used} is the $\used$ relation.
The pattern is then matched against the graph and the variables are bound to nodes. 
The filter condition predicates on the values of properties associated to the nodes.
Here the value of \texttt{data.Status} is expected to be one of the constants in the \texttt{classification} list.
This predicate selects all nodes with value \textit{at least} \texttt{Secret} in the ordered list.
The activity nodes that satisfy the conditions have their sensitivity set to 7.\footnote{A default value can be specified, i.e. for the cases where a \texttt{data} node has no \texttt{Status} property, or the property has no value.}
Rather than allowing arbitrary regular path expressions in the language, we expose specific traversal operators. 
One example is  \textit{descendantOf}, which returns all nodes reachable from a given start node.
An example of its use is the second rule above.
Rule evaluation binds variables \texttt{process} and \texttt{data} to activity and entity nodes $a$, $e$, respectively, such that $\used(a,e)$ holds and $e$ is any node that is reachable from node with id \texttt{d14} (a constant value).

Utility is the counterpart to sensitivity. It denotes the interest of the provenance owner in  ensuring that a node be \textit{retained} as part of the graph, as it represents important evidence which is not sensitive. 
Recall from our earlier example that grouping may remove non-selected nodes in order to preserve validity, a possibly undesirable side-effect.
The utility values associated to different nodes are used to quantify such loss of utility. 
Let $V_{ret} =  V \setminus V_{gr}$ be the set of nodes not intended for grouping, and $V'_{ret} \subset V_{ret}$ the nodes which were in fact retained after grouping.
The residual utility is simply
\begin{equation}
\mathit{RU}_V = \frac{\sum_{n \in V'_{ret}} u(n)}{ \sum_{n \in V_{ret}} u(n)} 
\label{eq:1}
\end{equation}
which is maximized for  $V'_{ret} = V_{ret}$.
Policy setters who experiment with different policy rules, i.e., using a test set of provenance graphs, may use $\mathit{RU}_V$ as a quantitative indicator of utility loss associated with a given policy and receiver.

%

 \vspace{-10pt}
\subsection{\provabs{} tool}

The Provenance Abstraction Model is implemented as part of a project involving confidentiality protection for provenance. 
The main purpose of the \provabs{} tool is to let a PS explore partial disclosure options, by experimenting with various policy settings and clearance level thresholds.
Users may load a graph in PROV-N format \cite{w3c-prov-n} and either specify a policy interactively, or load a pre-defined policy file.
%
The output consists of a graphical depiction of the graph, annotated with its sensitivity values (these are the coloured boxes in Fig. \ref{fig:sensitivities}), as well as the final abstract version of the graph.
The residual utility value (\ref{eq:1}) is also returned.
Provenance graphs are stored in the Neo4J graph database (\url{neo4j.org}).
Policy expressions are evaluated using a combination of the Neo4J Traverse API and Cypher queries.
\provabs{} and its documentation are publicly available.\footnote{\url{http://bit.ly/1dxg9X1}.}

\section{Summary}

In this paper we have presented a Provenance Abstraction Model (PAM) and its implementation, \provabs{}. PAM is based on a $\group$ operator, which replaces a set of nodes in a PROV graph with a new abstract node while preserving the validity of the graph. 
A simple notion of convexity of the set of nodes to be replaced ensures that the rewriting does not introduce new cycles.
Due to space limitations, the scope of this paper is limited to $\guEA$   graphs, which only include generation, usage relations on Activity and Entity nodes. 
A more comprehensive model, including its extension to Agents, can be found in our report \cite{Missier2013d}.
Encouraged by this initial study, we are now developing a more comprehensive model of abstraction that accounts for larger fragments of PROV --- a complex specification in its own right.

\end{document}